\documentclass[10pt,journal,compsoc]{IEEEtran}

\ifCLASSOPTIONcompsoc
  \usepackage[nocompress]{cite}
\else
  \usepackage{cite}
\fi
\usepackage{cite,graphicx,amsthm,amsmath,amssymb,mathrsfs,epsf}
\usepackage[usenames]{color}
\usepackage{multirow}
\usepackage{tabularx}
\usepackage{stfloats}
\usepackage{setspace}
\usepackage{bbold}
\usepackage{bm}
\usepackage{subcaption}
\usepackage{bigstrut,array,multirow,tabularx}
\usepackage{tikz}
\usepackage{xcolor}
\usepackage{pgfplots}
\usepackage{dsfont}
\usepackage{bbm}
\usepackage{graphics}
\usepackage{graphicx}
\usepackage{subcaption}
\usepackage[ruled, vlined, linesnumbered]{algorithm2e}
\usepackage{soul}
\usepackage{booktabs}

\newtheorem{definition}{Definition}
\newtheorem{theorem}{Theorem}

\newtheorem{example}{Example}

\DeclareMathOperator{\E}{\mathds{E}}

\newcommand{\argmax}{\mathop{\mathrm{arg\,max}}}

\begin{document}

\title{
	\textsf{IM-META}: Influence Maximization Using Node Metadata in Networks With Unknown Topology
}

\author {
       	 Cong Tran, Won-Yong Shin,~\IEEEmembership{Senior Member,~IEEE}, and Andreas Spitz
\IEEEcompsocitemizethanks{\IEEEcompsocthanksitem C. Tran was with the Department of Computer Science and Engineering, Dankook University, Yongin 16890, Republic of Korea, and also with the Machine Intelligence \& Data Science Laboratory, Yonsei University, Seoul 03722, Republic of Korea. He is with the Faculty of Information Technology, Posts and Telecommunications Institute of Technology, Hanoi 100000, Vietnam.\protect\\
E-mail: congtt@ptit.edu.vn.
\IEEEcompsocthanksitem W.-Y. Shin is with the School of Mathematics and Computing (Computational Science and Engineering), Yonsei University, Seoul 03722, Republic of Korea, and also with the Graduate School of Artificial Intelligence, Pohang University of Science and Technology (POSTECH), Pohang 37673, Republic of Korea.\protect\\
E-mail: wy.shin@yonsei.ac.kr.
\IEEEcompsocthanksitem A. Spitz is with the Department of Computer and Information Science, University of Konstanz, Konstanz 78457, Germany.\protect\\
E-mail: andreas.spitz@uni-konstanz.de.\\
(Corresponding author: Won-Yong Shin.)}}%
\markboth{}
{}

\IEEEtitleabstractindextext{
\begin{abstract}
Since the structure of complex networks is often {\em unknown}, we may identify the most influential seed nodes by exploring only a part of the underlying network, given a small budget for node queries. We propose \textsf{IM-META}, a solution to influence maximization (IM) in networks with unknown topology by retrieving information from queries and node metadata. Since using such metadata is not without risk due to the noisy nature of metadata and uncertainties in connectivity inference, we formulate a new IM problem that aims to find both seed nodes and queried nodes. In \textsf{IM-META}, we develop an effective method that {\em iteratively} performs three steps: 1) we learn the relationship between collected metadata and edges via a {\em Siamese neural network}, 2) we select a number of inferred confident edges to construct a {\em reinforced} graph, and 3) we identify the next node to query by maximizing the inferred influence spread using our {\em topology-aware} ranking strategy. Through experimental evaluation of \textsf{IM-META} on four real-world datasets, we demonstrate a) the speed of network exploration via node queries, b) the effectiveness of each module, c) the superiority over benchmark methods, d) the robustness to more difficult settings, e) the hyperparameter sensitivity, and f) the scalability.

\end{abstract}

}
\maketitle

\IEEEdisplaynotcompsoctitleabstractindextext

%
\IEEEpeerreviewmaketitle

\section{Introduction}

\subsection{Background and Motivation}

\IEEEPARstart{S}{ocial} networks that encode the relationships and interactions of individuals play a vital role as a medium for facilitating the sharing of ideas, thoughts, and information through the building of virtual networks and communities. Well-known applications of such social networks include viral marketing---for example, a company would like to hire the most influential opinion leaders to advertise a new product. To this end, the so-called {\em influence maximization (IM)} problem that aims at identifying a set of seed nodes, leading to the maximum influence spread under given influence diffusion models, was introduced as a formalization \cite{IMorigin}. In addition to (viral) marketing in social networks, efficient solutions to the IM problem stand to benefit applications in numerous domains such as microfinance \cite{c2} and healthcare \cite{HEALER}.

In the original formulation of the IM problem, the topological structure of the underlying network in question is assumed to be fully known \cite{IMorigin}. In practice and despite the growing popularity of studies on IM, however, this assumption tends to be overly simplistic as the underlying network cannot be fully observed in many real-world applications of network analyses  \cite{c1,HIV}. Although one could, in principle, invest further efforts into uncovering the complete network prior to the application of existing IM algorithms, the accumulation of a complete network structure is often prohibitively expensive, labor-intensive, or entirely impractical~\cite{c1}. Even with the increasing availability of (social) network data from social media platforms, this source of information is not without limitations, not least due to privacy concerns \cite{datacollection}. 
Consequently, when working with graph data, one should assume that none or only a part of the network structure is available in practice  \cite{metacode,kronem,deepnc,kromfac}.

When confronted with the need to perform IM on networks that have such an unknown or partially known topology, a naturally arising question is therefore how one can identify a set of seed nodes, necessitating the network structure. Recent studies have addressed this question by allowing a small budget for {\em node queries} to surgically discover parts of the underlying network, while simultaneously aiming to identify a set of seed nodes from the explored subgraph that is expected to be as influential as the globally optimal seed set \cite{c3,c4,EIM,EIM2,EIM3,EIM4,HEALER}. However, node metadata ({\it i.e.}, node features), which can often be retrieved from various sources such as users’ created content over given (social) networks \cite{coldstart} or even be inferred from users' individual behavior \cite{inferattrib}, are not fully considered in this problem despite the fact that they are more readily available than the relationship between individuals. Such metadata may then be leveraged to identify relations that are likely to exist. For example, in co-authorship networks, one might take advantage of the keywords in each author’s publications to infer a topical relationship between authors. In the IM problem for HIV awareness in homeless youth \cite{HIV}, metadata such as the individual's name, age, and familial background among the homeless demographic  as well as individuals can be readily gathered due to the mandatory registration process at welfare centers, which is necessitated for the disbursement of monthly allowances. Motivated by the observation that the collection of metadata for nodes is cost-effective, we propose a solution for real-world applications of IM in incomplete (social) network settings where the underlying network $G$ is {\em initially unknown}.


\subsection{Main Contributions}

In this study, we present \textsf{IM-META}, an end-to-end framework that solves the IM problem in a network $G$ with minimal information about $G$, using node queries aided by {\em node metadata} in such {\it attributed} networks. To this end, we aim to retrieve information from {\em jointly} discovering a sequence of node queries and a set of nodes that maximize the influence spread by formulating an entirely new optimization problem that jointly finds the optimal discovered subgraph and seed set in the sense of maximizing an {\em estimate} of the expected influence spread.
Naturally, solving this optimization problem poses several practical challenges, namely 1) the difficulty of handling noisy node metadata, 2) the scarcity of training labels due to using a small budget for node queries, and 3) the intractability of jointly identifying two optimal subsets ({\it i.e.}, the subgraph and the seed nodes) and dynamically updating an estimate of the expected influence spread for every query step. The combination of the first two challenges is particularly difficult since the redundancy of observable yet inherently noisy node features in combination with the small query budget is likely to entail the curse of dimensionality that threatens to degrade the inference accuracy. 

To address these challenges, we propose a new design principle with \textsf{IM-META}, which utilizes a two-phase system consisting of a network discovery phase (NDP) and a seed selection phase (SSP). First, network discovery is performed in three iteratively repeated steps. Each iteration starts with the first step, NDP1, in which we infer a set of existence probabilities for yet unexplored edges using both the node metadata and the currently explored subgraph. We utilize a {\em minimally supervised} learning architecture that is based on a {\em Siamese neural network} model capable of learning the relationship between collected metadata and edge prediction by using the explored small-scale subgraph as training data. By learning embeddings that accurately capture the {\em homophily} effect of information diffusion networks ({\it i.e.}, a node’s tendency to associate with similar other nodes \cite{homophily}), the Siamese neural network inductively embeds new input data ({\it i.e.}, unexplored nodes) whose connectivity information to the training data ({\it i.e.}, explored nodes) is unknown. The learned embedding then acts as a feature extraction module (or a dimensionality reduction module) that captures the latent representation of the collected data. In NDP2, we generate a {\em reinforced} weighted graph by combining the edge probabilities inferred in NDP1 with the current explored subgraph. To this end, we select a small number of {\em {\it confident} edges} based on the intuition that 1) using a large portion of uncertain edges leads to an accumulation of noise in the subsequent steps and phase, and that 2) this edge selection also reduces the computational complexity. In the third step, NDP3, we  select one queried node for network discovery. In this step, one of our central contributions is a novel \emph{topology-aware} ranking strategy for query node selection that balances the degree centrality and the geodesic distance to potential seed nodes. As a result, each node that is selected with this ranking strategy has a high expected influence spread. After the completion of the NDP, we select a set of seed nodes in the SSP, similar to the standard IM problem~\cite{IMorigin,celf,IMM,TIM}. 

Our contributions are four-fold:
\begin{itemize}
\item We introduce \textsf{IM-META}, a novel end-to-end framework that solves the IM problem in topologically unknown networks;

\item We formalize IM on unknown networks as an optimization problem that aims to find both a set of seed nodes and another set of nodes to query by making use of node metadata;

\item We design an effective approach for network discovery that iteratively performs three steps, namely network inference, reinforced weighted graph generation, and query node selection;

\item Through extensive experiments on real-world datasets with node features across several domains, we validate (a) the fast subgraph exploration, (b) the achievability of the upper bound performance via querying only 5\% of nodes, (c) the effectiveness of Siamese neural network-based network inference and topology-aware ranking-based query node selection modules, (d) the superiority of  \textsf{IM-META} over benchmark IM methods, (e) the robustness of our method to incomplete node metadata, and (f) scalability.
\end{itemize}

To the best of our knowledge, this is the first proposed solution to the IM problem in attributed networks with unknown topology. 

\subsection{Organization}

The remainder of this paper is organized as follows. In Section \ref{sec:2}, we present prior studies related to our work. In Section \ref{sec:3}, we explain the methodology of our study, including the problem formulation and an overview of
our \textsf{IM-META} method. Section \ref{sec:4} describes technical details of the proposed method and analyzes its computational
complexity. Experimental results are shown in Section \ref{sec:5}. Finally, we provide a summary and concluding remarks in Section \ref{sec:6}.

\section{Related Work}\label{sec:2}

\begin{table*}[t]
\centering
\caption{Summary of IM methods for topologically unknown networks using communities (Com.), influenced cascades (Inf. Cas.), similar graphs (Sim. Gra.), and domain expert knowledge (Dom. Exp.).}
\label{tab:literature}
\centering
\begin{tabular}{l c c c c c c}
\toprule
\multirow{2}{*}{Methods}          & \multicolumn{6}{c}{Required side information}                   \\ \cmidrule(l){2-7} 
                                 & Degree  & Com.            & Inf. Cas.            & Sim. Gra.            & Dom. Exp.     & Metadata       \\ \cmidrule(r){1-7}
IMUG \cite{heuristic} & \checkmark  &     &            &       &          &                                         \\  
ARISEN \cite{EIM}                 &   & \checkmark &    &          &               &         \\
NeuGreedy \cite{cascade}             &        &    &    \checkmark &       &   &  \\ 
DQN \cite{EIM2}, CLAIM \cite{EIM3}, RL4IM \cite{EIM4}       &                &                  &  &  \checkmark      &  & \\ 
DOSIM \cite{DOSIM}, HEALER \cite{HEALER}&   &    &             &          &      \checkmark     &       \checkmark                                    \\ 
\textsf{IM-META}                       &        &          &   &        &  & \checkmark\\ 
\bottomrule
\end{tabular}
\end{table*}

The method that we proposed in this study is related to the following four broad research lines.

{\bf IM in topologically known networks.} Kempe et al. \cite{IMorigin} originally introduced the problem formulation of IM when the network structure is initially known and proposed a greedy approach to solve it. Besides heuristic solutions scalable to large-scale networks \cite{degreediscount,IRIE,SAMIR,TANG}, subsequent studies focused mostly on reducing the computational complexity of greedy algorithms, such as CELF/CELF++ \cite{celf,WWW2}, TIM/TIM+ \cite{TIM}, IMM \cite{IMM}, SSA/D-SSA \cite{SSA}, RS \cite{RS}, and CFDI \cite{comIM}, while providing approximate solutions with performance guarantees.

{\bf IM in topologically unknown networks.} More recently, researchers have paid attention to a more challenging IM problem, in which the structure of the underlying network is initially unknown. Following the  concept of active learning for classification \cite{active,activeTPAMI}, dynamic IM over a series of rounds in which information on edges is collected after each round has been studied \cite{heuristic}. To aid the network discovery process, exploratory IM (EIM) methods exploit topological information such as the community structure \cite{EIM}. Recently, EIM methods tend to rely on complementary data such as a set of similar graphs \cite{EIM2,EIM3,EIM4}. However, such data are scarcely available under the topology-unaware setting of many real-world networks.  As follow-up studies on EIM, CHANGE \cite{c3} utilized the friendship paradox, while DQN \cite{EIM2}, CLAIM \cite{EIM3}, and RL4IM \cite{EIM4} learned patterns from a set of similar graphs to effectively select a set of query nodes. 
Another attempt to solve the IM problem in unknown networks involves learning an influence function based on observed cascades \cite{cascade}. Although some attempts have been made to exploit node metadata \cite{HEALER,DOSIM}, they rely on handcrafted metadata, thereby incurring significant annotation efforts by domain experts that lack scalability since such work is infeasible on large network datasets. Moreover, a theoretical analysis on the performance of IM was carried out in settings where a subgraph is retrieved via random node sampling  \cite{pvim,seedcostly}. Table~\ref{tab:literature} summarizes the aforementioned IM methods utilizing different types of side information in topologically unknown networks.

{\bf IM in temporal/dynamic networks.} Recently, there have been several studies that explore the IM problem for temporal and dynamic networks with different settings. Assuming that the past and current snapshots of the underlying network are known, the authors attempted to forecast its evolution using non-negative matrix factorization and graph neural networks \cite{tempIM2}. In another setting, under the assumption that the network changes over time, it was studied that such changes can be obtained by probing nodes and updating the topology \cite{tempIM3}. 
In a similar probing context, a divide-and-conquer strategy was used in \cite{tempIM1} to minimize the variance between the observed topology and the ground truth network across diverse representative communities. 

{\bf Link prediction using metadata.} When the network structure is partially observable, a matrix factorization-aided approach \cite{matrixfac} was proposed to leverage node features for link prediction. Our study is related to the {\em inductive} link prediction problem, the so-called cold-start link prediction, in which no initial topological information is available. To achieve satisfactory prediction performance, deep learning-based techniques such as G2G \cite{g2g} and DEAL \cite{deal} have been proposed, where they use a deep encoder that embeds each node in a given network as Gaussian distributions and aligns both structure and feature embeddings, respectively.

{\bf Discussion.} Despite numerous contributions to IM in unknown networks, our  method is the first to exploit the collected {\em node metadata} to automatically aid the process of discovering influential seed nodes. Unlike prior studies that perform inductive link prediction in a supervised manner ({\it e.g.}, graph neural network-based methods), our solution tackles the lack of explored edges that can be used as labels in training when only a small portion of the underlying network is observable via {\it node queries}.

\section{IM in Networks With Unknown Topology}\label{sec:3}

As a basis for the proposed \textsf{IM-META} method in Section \ref{sec:4}, we first describe our network and influence diffusion models with basic assumptions. We then formulate a new problem using these models. 

\subsection{System Model}\label{sec:3a}

\subsubsection{Network Model}

Let us denote an underlying network with a-priori unknown topology as $G = (\mathcal{V}, \mathcal{E})$, where $\mathcal{V}$ is the set of $n$ nodes and $\mathcal{E}$ is the set of $m$ edges. We assume $G$ to be an undirected unweighted network without self-loops or repeated edges  and to have collectible {\em node metadata} ({\it i.e.}, node features) $\mathcal{X}$ for all $n$ nodes.
The metadata in $\mathcal{X}$ associated with each node $v_i \in \mathcal{V}$ are assumed to be given in a vector form of $d$ features, denoted as ${\bf x}_i \in \mathbb{R}^{d}$.

\begin{figure*}[t]
    \begin{center}
            \includegraphics[width=\linewidth]{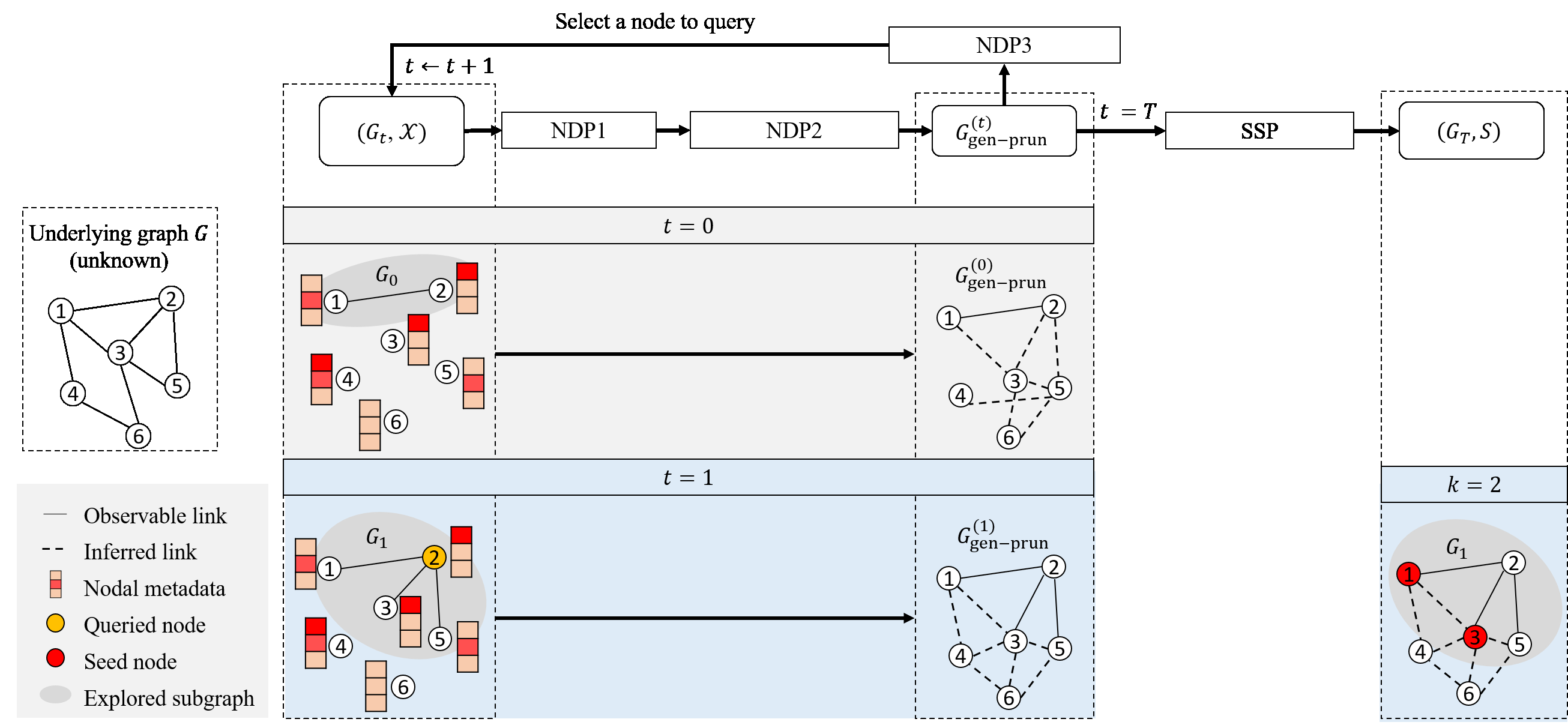}
            \caption{A schematic overview of our proposed  \textsf{IM-META} method for $T=1$ node query and $k=2$ seed nodes. Here, after adding seven inferred edges to obtain $G^{(0)}_\text{gen-prun}$, node $v_2$ (a filled-in yellow circle) is selected. Then, the seed set of two nodes $v_1$ and $v_3$ (filled-in red circles) is chosen from $\mathcal{V}_1$ in the subgraph $G_1$.}
            \label{fig:exampleoverall}
    \end{center}
\end{figure*}

Following the definition by Kamarthi et al. \cite{EIM2}, an induced subgraph $G_0$ is assumed to be observable at a very early stage. When $G_0$ is not initially available, such a subgraph can be obtained by querying a few nodes at random. To retrieve additional structural information from the underlying network $G$, we are allowed to perform $T$ node queries. Specifically, when we query a single node, we are capable of discovering neighbors of the queried node. The observable subgraph can then be expanded accordingly, which essentially follows the same setting as prior work in the field \cite{EIM,EIM2,heuristic}. Let $G_t = (\mathcal{V}_t, \mathcal{E}_t)$ be the subgraph discovered after $t$ queries, where $\mathcal{V}_t$ and $\mathcal{E}_t$ are the set of nodes and the set of edges, respectively, in the subgraph $G_t$. During the $(t+1)$-th node query, we choose a node $v_t$ from $G_t$, which enables us to observe an expanded graph $G_{t+1} = (\mathcal{V}_t \cup \mathcal{N}_G(v_t), \mathcal{E}_t \cup \mathcal{E}(\mathcal{N}_G(v_t), {v_t}))$, where $\mathcal{N}_G(v_t)$ indicates a set of neighbors of node $v_t$ in $G$ and $\mathcal{E}(\mathcal{N}_G(v_t),v_t)$ is a set of all edges to which each node in $\mathcal{N}_G(v_t)$ and node $v_t$  are incident. An example of this query process is illustrated in Fig. \ref{fig:exampleoverall}, where an initially induced subgraph $G_0$ consists of two nodes $v_1$ and $v_2$, and we observe another graph $G_1$ consisting of four nodes ({\it i.e.}, $v_1$, $v_2$, $v_3$, and $v_5$) after querying node $v_2$. The finally  discovered subgraph $G_T = (\mathcal{V}_T, \mathcal{E}_T)$ is used as input of the IM process, where a set of seed nodes ({\it i.e.}, a subset of observable nodes) of size $k$, $S\subseteq \mathcal{V}_T$, is chosen to diffuse the influence.\footnote{Due to the limited availability of the network structure, it is typical to select seed nodes only among the retrieved nodes. We refer to \cite{EIM,EIM2,pvim} for more details.}

\subsubsection{Influence Diffusion Model}

We consider two models of influence diffusion, namely the independent cascade (IC) and weighted cascade (WC) models \cite{degreediscount,pvim,IMorigin}. In the IC model, each edge is associated with a uniform diffusion probability. In the WC model, this probability is modulated by the in-degree of the node being influenced.

\subsection{Problem Formulation}\label{sec:3b}

In this subsection, we formulate a new IM problem for attributed networks whose topology is unknown and node features are available.
In a given diffusion model, we define a function $I(S)$ : $2^{\mathcal{V}_T} \rightarrow [0, n]$ that maps a seed set $S$ to the number of influenced nodes in the underlying graph $G$ at the end of the diffusion process, which  is a random variable since the same set $S$ results in different network realizations. 
In our study, based on the concept of active learning \cite{active}, we aim to maximize the expected influence spread $\sigma(S)=\E(I(S))$ by {\em jointly} finding a sequence of queries $(v_0,v_1, \cdots,v_{T-1})$ that leads to the finally discovered subgraph $G_T$, as well as a seed set $S \subseteq \mathcal{V}_T$, given a query budget $T$ and a seed budget $k$. This is also referred to as an {\em IM problem with exploration} in the literature \cite{EIM,EIM2}.
In contrast to the conventional setting of IM~\cite{IMorigin}, since the influence spread $\sigma(\cdot)$ cannot be computed accurately from $G$ due to the uncertainty of the unexplored part of $G$, our problem is ill-defined. Thus, we need to compute an estimate of  $\sigma(\cdot)$ based on the available information.

To this end, we first gradually infer the connections between two nodes in the unexplored part based on both the explored graph and the set of node features, $\mathcal{X}$, by repeatedly querying a single node in each step, which corresponds to {\em network inference} in our study. An edge ($u,v$) $\notin \mathcal{E}_t$ that is incident to nodes $u$ and $v$ that have not yet been queried is referred to as an {\em uncertain} edge. We note that there are three types of such uncertain edges, including i) the edges connecting two nodes in $\mathcal{V}_t$, ii) the
edges connecting one node in $\mathcal{V}_t$ and another node in $\bar{\mathcal{V}}_t$,
and iii) the edges connecting two nodes in $\bar{\mathcal{V}}_t$, where $\bar{\mathcal{V}}_t$ is the set of unexplored nodes at the $t$-th query step such that $\mathcal{V}=\mathcal{V}_t \cup \bar{\mathcal{V}}_t$ for two disjoint sets $\mathcal{V}_t$ and $\bar{\mathcal{V}}_t$. Intuitively, during the $t$-th node query, we are capable of learning a {\em generative model of edges}, which approximates the distribution of all connections in $G$ given the set of node features in $\mathcal{V}_t$ and the subgraph $G_{t}$ discovered after $t$ queries as training samples. Using this learned model, we then predict the probability $\theta^{(t)}_{uv} \in [0,1]$ that an uncertain edge $(u,v)$ is present, the so-called edge probability. The subgraph $G_t$ and the set of $\theta_{uv}^{(t)}$'s for all uncertain edges are fed into the generator function of a weighted graph, denoted as $G_\text{gen}^{(t)}$, having all $n$ nodes. By integrating the connectivity information of both observable and predicted edges via the generator function, we are capable of creating  $G^{(t)}_\text{gen}$ whose adjacency matrix ${\bf A}^{(t)}_\text{gen}$ is given by
\begin{equation}
\label{eq:impute}
    a_{ij}= 
\begin{cases}
   1, & \text{\upshape if } (i,j) \in \mathcal{E}_t\\
   0, & \text{\upshape if } (i,j) \notin \mathcal{E}_t \text{ \upshape and } i,j \notin \mathcal{Q}_t  \\ 
   \theta^{(t)}_{ij},              & \text{\upshape otherwise},
\end{cases}
\end{equation}
where $\mathcal{Q}_t$ is the set of all queried nodes until $t$ queries and  ${a}_{ij}$ denotes the $(i,j)$-th element of ${\bf A}^{(t)}_\text{gen}$  for $i\neq j$.

Next, using the generated weighted graph $G_\text{gen}^{(t)}$, we turn to explaining how to compute an estimate of  the expected influence spread of $\sigma(\cdot)$, denoted as $\hat{\sigma}(\cdot)$. Let $E_{u,v}$ denote the event that there exists an edge between nodes $u$ and $v$. Then, it follows that $p(E_{u,v}) = \theta^{(t)}_{uv}$. Let us denote $A_{u,v}$ as the event that a node $u$ successfully activates another node $v$ in the diffusion process. From the fact that the event $A_{u,v}$ is valid only when an incident edge $(u,v)$ is present, the activation probability that models the strength of influence that node $u$ has on another node $v$ is expressed as the conditional probability $p(A_{u,v}|E_{u,v})$. Thus, we can compute the joint probability distribution $p(E_{u,v},A_{u,v})$ for each edge in $\bar{\mathcal{E}}_t$ using Bayes rule. Now, it is possible to  compute an estimate  $\hat{\sigma}(S)$ given a set of seed nodes, $S\subseteq \mathcal{V}_T$, by taking the expectation over all edges $(u,v)$ in $G_\text{gen}^{(t)}$  via Monte-Carlo simulations. Finally, we are ready to formulate our new problem that aims at finding the optimal discovered subgraph and seed set $(G_T^*,S^*)$ in the sense of maximizing $\hat{\sigma}(S)$ as follows:
\begin{equation}
\label{eq:probform}
(G_T^*,S^*) = \argmax_{G_T, S \subseteq \mathcal{V}_T,|S|=k}\hat{\sigma}(S). \\
\end{equation}

Intuitively, the discovered subgraph $G^*_T$ should be optimally found in the sense of containing as many optimal seed nodes as possible. Thus, it would be optimal in case all optimal seed nodes $S^*$ are included in  the subgraph $G_T^*$ that has been discovered before the seed set selection, {\it i.e.}, $S^*\subseteq \mathcal{V}_T^*$  for the set of vertices $\mathcal{V}_T^*$ in $G_T^*$. However, we note that the inference accuracy of $\hat{\sigma}(S)$ plays an important role in determining the influence  ({\it i.e.}, the number of activated nodes) in the network $G$ since $\hat{\sigma}(\cdot)$ is just an estimate of the actual expected spread.   Thus, the designed method should consolidate an accurate estimate of ${\sigma}(S)$.

In our study, we tackle two practical challenges as follows: 1) we consider a realistic setting in which the number of queries that can be used to obtain training data for the generative model is strictly limited and 2)  the problem of solving (\ref{eq:probform}) is NP-hard with an exponential complexity in searching for the two sets $G_T$ and $S$.\footnote{Since our problem maps to the original IM problem when $G_T	\equiv G$, the proof of NP-hardness similarly follows that of the original  problem \cite{IMorigin}.} This motivates us to present an {\em accurate} and {\em low-complexity} solution to this problem  in (\ref{eq:probform}).

\section{\textsf{IM-META} Method}\label{sec:4}

In this section, we present \textsf{IM-META}, an end-to-end solution to the problem of IM in a topologically unknown attributed network. Our main idea is to solve two subproblems where we first update $G_T$ while fixing an estimated $S^*$, then optimizing $S$ with fixed $G_T$. In particular, due to the interdependence of the seed set $S$ on the explored subgraph $G_T$, where $S$ is chosen from vertices within $G_T$, we can represent $S$ as a function of $G_T$ and vice versa, denoted as $S = f(G_T)$ and $G_T = g(S)$. Consequently, our objectives encompass addressing the following two iterative subproblems: 1) when presented with $G_T$, our aim is to identify the optimal $S$; and 2) when provided with an estimated $S^*$, we proceed to optimize $G_T$.
 The overall procedure of our method is described in Algorithm 1. The algorithm starts with the NDP consisting of the following three functions: 1) \textsf{NetInference} for network inference using metadata  (refer to line 3); 2) \textsf{GraphGen}  for reinforced weighted graph generation (refer to line 4); and 3) \textsf{QNodeSelect} for query node selection  (refer to line 5). After querying $T$ nodes by iteratively invoking the above three functions, we turn to the SSP along with a modified greedy IM function, dubbed \textsf{ModifiedGreedy}, which is invoked to obtain the set of influential seed nodes, $S$, from  the reinforced weighted graph $G^{(T)}_\text{gen-prun}$ (refer to line 9).

\begin{example}
A schematic overview of our \textsf{IM-META} method is visualized in Fig.~\ref{fig:exampleoverall}, where we show one round of exploration from an unknown attributed network  $G$ consisting of six nodes. As output of the network inference (NDP1) and the reinforced weighted graph generation (NDP2), we acquire a network $G^{(0)}_\text{gen-prun}$ by adding seven inferred edges. Based on $G^{(0)}_\text{gen-prun}$, node $v_2$ (a filled-in yellow circle) is selected via query node selection (NDP3). Afterwards, we similarly obtain $G^{(1)}_\text{gen-prun}$, which completes the NDP. Under a given budget of $k=2$ seed nodes, the seed set of two nodes $v_1$ and $v_3$ (filled-in red circles) is chosen from $\mathcal{V}_1$ in the subgraph $G_1$ since the seed set maximizes the influence spread over the reinforced graph $G^{(1)}_\text{gen-prun}$ by employing our own greedy IM algorithm.
\end{example}

\begin{algorithm}[t]
\DontPrintSemicolon
\KwIn{$G_0,\mathcal{X},T,k$}
\KwOut{$G_T,S$}
\SetKwBlock{Begin}{function}{end function}
\Begin($\textsf{IM-META}$)
{
  \For{$t$ {\upshape \bf from} $0$ {\upshape \bf to} $T-1$}
  {
  $\Theta^{(t)} \leftarrow$ \textsf{NetInference}($\mathcal{X},G_t$) (Section 4.1.1)\;
  $G^{(t)}_\text{gen-prun} \leftarrow$ \textsf{GraphGen}($\Theta^{(t)}, G_t$) (Section 4.1.2)\;
  $v_t \leftarrow$ \textsf{QNodeSelect}($G^{(t)}_\text{gen-prun}, k$) (Section 4.1.3)\;
  $G_{t+1} \leftarrow (\mathcal{V}_t \cup \mathcal{N}_G(v_t), \mathcal{E}_t \cup \mathcal{E}(\mathcal{N}_G(v_t), {v_t}))$\;
  }
    $\Theta^{(T)} \leftarrow$ \textsf{NetInference}($\mathcal{X},G_T$)\;
  $G^{(T)}_\text{gen-prun} \leftarrow$ \textsf{GraphGen}($\Theta^{(T)},G_T$)\;
  $S \leftarrow$ \textsf{ModifiedGreedy}($G^{(T)}_\text{gen-prun},k$) (Section 4.2)\;
  \Return{$G_T,S$}
}
\caption{\textsf{IM-META}}\label{al1}

\end{algorithm}

\subsection{Network Discovery}\label{sec:4a}

The technical details of three steps in the NDP are described in the following.

\subsubsection{Network Inference Using Node Metadata (NDP1)}\label{sec:4a1}

We elaborate on how to infer a set of uncertain edge probabilities, denoted as $\Theta^{(t)}$, at each query step $t \in \{1,\cdots,T\}$ using node metadata. Such network inference, treated as inductive link prediction, entails several practical challenges. First, not all collected features in the set $\mathcal{X}$ are useful for the inference process due to redundancy and computational inefficiency. Second, a limited budget for node queries results in a small-size explored subgraph. Since none of the incident edges of each unexplored node in the subgraph are visible, we need to {\em inductively} infer $\Theta^{(t)}$ based soly on the node metadata $\mathcal{X}$.

We tackle the aforementioned challenges by adopting a {\em Siamese neural network} model, which was originally introduced in image matching \cite{Siamese}. In our work, we take advantage of the ``homophily principle" in social networks, the tendency of an individual node to associate with similar others \cite{homophily}, and adopt the Siamese neural network model with the primary role of dimensionality reduction. To this end, we learn embeddings that accurately capture the {\em homophily} effect of information diffusion networks. Since a trained Siamese neural network inductively maps new input data ({\it i.e.}, unexplored nodes) whose connectivity information to the training data  ({\it i.e.}, explored nodes) is unknown, it can easily be applied to our problem. Our Siamese neural network consists of twin multilayer perceptron (MLP) networks to which two feature vectors ${\bf x}_u$ and ${\bf x}_v$ associated with nodes $u$ and $v$, respectively, are fed separately. The MLPs act as encoders that compute embedding vectors, denoted as ${\bf e}_u$ and ${\bf e}_v$ for nodes $u$ and $v$, respectively. Finally, similarly as in \cite{hadamard}, we compute the Hadamard product of two embedding vectors and then utilize a sigmoid output layer to infer $\theta_{uv}^{(t)}$ from the product. That is, $\theta_{uv}^{(t)} = \text{sigmoid}({\bf e}_u \odot {\bf e}_v)$, where $\odot$ denotes the Hadamard product. 

\subsubsection{Reinforced Weighted Graph Generation (NDP2)}\label{sec:4a2}

We now explain how to seamlessly generate a reinforced weighted graph. From the step NDP1 ({\it i.e.}, network inference), we obtain edge probabilities $\Theta^{(t)}$ at each query step $t$. However, a large portion of these uncertain edges do not truly exist since real-world graphs are usually sparse. Thus, using non-existent edges certainly leads to an {\em accumulation of noise} in the subsequent processes. This motivates us to select edges that are  predicted with high confidence to reduce noise while simultaneously reducing the computational complexity. To this end, we aim to select {\em {\it confident}} edges whose inferred probabilities $\Theta^{(t)}$ are greater than a given threshold as follows.

\begin{definition}\label{def:2}
Suppose that the set $\Theta^{(t)}$ indicating the probability of uncertain edges is inferred at the $t$-th query. Then, an edge $(u,v) \in \Theta^{(t)}$ is called a {\it confident} edge if $\theta_{uv}^{(t)} \geq \epsilon$, where $\epsilon > 0$ is a predefined threshold. 
\end{definition}

Let $H\in\{0,1,\cdots,|\Theta^{(t)}|\}$ denote the cardinality of the set of selected {\it confident} edges. After selecting the top $H$ {\it confident} edges in the ranked list of $\Theta^{(t)}$ in descending order, we transform the weighted graph $G_\text{gen}^{(t)}$ into a {\em reinforced} weighted graph  $G_\text{gen-prun}^{(t)}$  via our {\em edge prunning} strategy. Then, we recompute the diffusion probabilities for all edges in $G^{(t)}_\text{gen-prun}$. Specifically, we compute the joint probability distribution based on each diffusion model as follows. For the IC model, we have $\text{Pr}(A_{u,v}|E_{u,v}) = p_{uv}\sim\mathcal{U}(0,1)$.
Thus, from Bayes' rule, it follows that $\text{Pr}(E_{u,v}A_{u,v}) = \theta_{uv}^{(t)}p_{uv}$. For the WC model, we have $\text{Pr}(A_{u,v}|E_{u,v}) = \frac{1}{d_v}$, which is however unknown since $d_v$ representing the degree of node $v$ can only be computed from the fully observed network structure. To overcome this problem, we estimate the node degree at step $t$ as $\hat{d}_v^{(t)}:=\sum_{u \in \mathcal{N}_v} \theta_{uv}^{(t)}$, where $\mathcal{N}_v$ is the set of all neighbors of $v$ in $G_\text{gen-prun}^{(t)}$ and $\theta_{uv}^{(t)} = 1$ if $(u,v) \in \mathcal{E}_t$. Then, we have $\text{Pr}(E_{u,v}A_{u,v}) =  \frac{1}{\hat{d}_v^{(t)}}\theta_{uv}^{(t)}$ for the WC model.

\begin{figure}[t]
    \begin{center}
            \includegraphics[width=\linewidth]{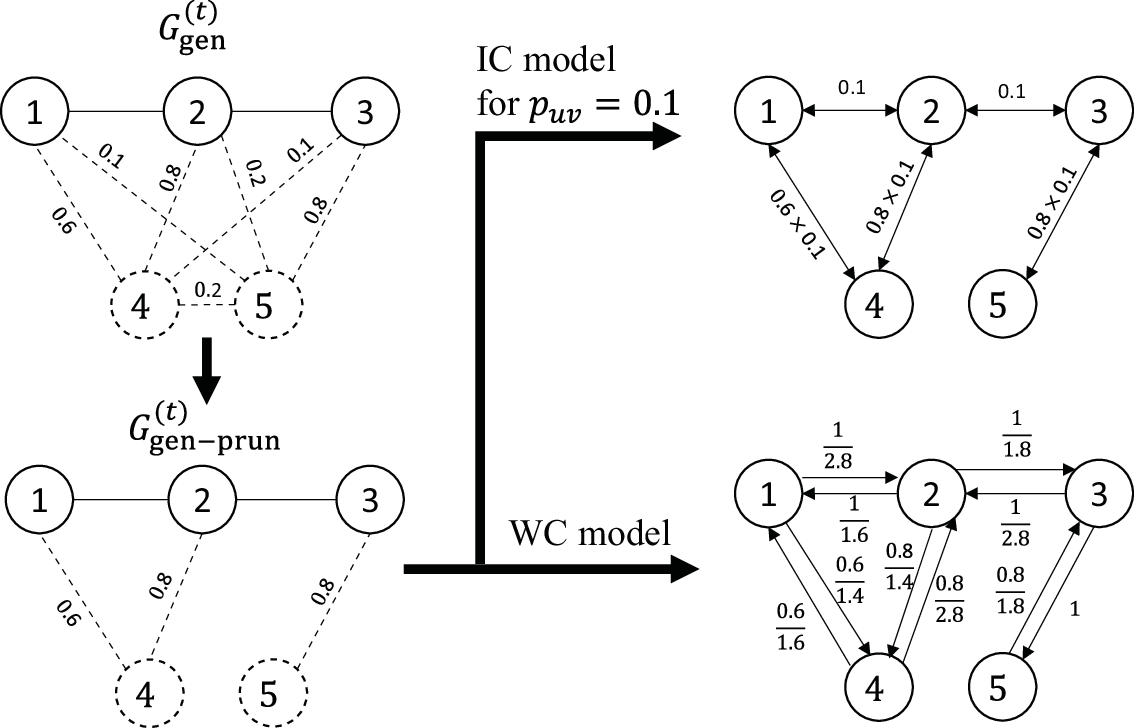}
            \caption{An illustration of reinforced weighted graph generation for $\epsilon=0.5$ under the two diffusion models. Here, the value on each dashed line indicates $\theta_{uv}^{(t)} \in \Theta^{(t)}$, while the value on each solid line represents the diffusion probability in the reinforced weighted graph $G_\text{gen-prun}^{(t)}$.}
            \label{fig:example3}
    \end{center}
\end{figure}

\begin{example}
An example of the reinforced weighted graph generation can be seen in Fig.~\ref{fig:example3}, where $\epsilon$ = 0.5 and two reinforced weighted graphs $G_\text{gen-prun}^{(t)}$ are constructed based on both IC and WC models. For the IC model, assuming that $p_{uv} = 0.1$, the diffusion probability in $G_\text{gen-prun}^{(t)}$ can be simply computed as in the right-top of Fig.~\ref{fig:example3}. For the WC model, we compute the diffusion probabilities of associated edges in $G_\text{gen-prun}^{(t)}$ by using an estimate of the node degree, $\hat{d}_v$ (see the right-bottom of Fig.~\ref{fig:example3}). For example, $\hat{d}_1^{(t)}$ is given by $1+\theta_{41}^{(t)}$ = 1 + 0.6 = 1.6; thus, the diffusion probabilities associated with two edges $(v_2,v_1)$ and $(v_4,v_1)$ are $\frac{1}{1.6}$ and $\frac{0.6}{1.6}$, respectively.
\end{example}

\subsubsection{Query Node Selection (NDP3)}\label{sec:4a3}

As the last step of network discovery, we describe our query node selection module. As in prior studies~\cite{EIM, EIM2}, it is natural to select a query node that has a high degree centrality since it enables the resulting explored subgraph to grow fast. In our work, selecting a high degree node to query would be beneficial in providing more samples to train our Siamese neural network. Instead, we pay attention  to the {\em residual degree} of a node at each query step $t$, defined as $r_u^{(t)} = \hat{d}_u^{(t)} - d_u^{(t)}$, rather than the estimated degree $\hat{d}_u^{(t)}$  itself, where $d_u^{(t)}$ denotes the degree of an observable node at  the $t$-th query step. This is because obviously,  the residual degree $r_u$ represents a better capability to explore more nodes when the inference accuracy is sufficiently high since the observable nodes in the explored subgraph no longer contribute to its growth.

\begin{figure}[t]
    \begin{center}
            \includegraphics[width=0.55\linewidth]{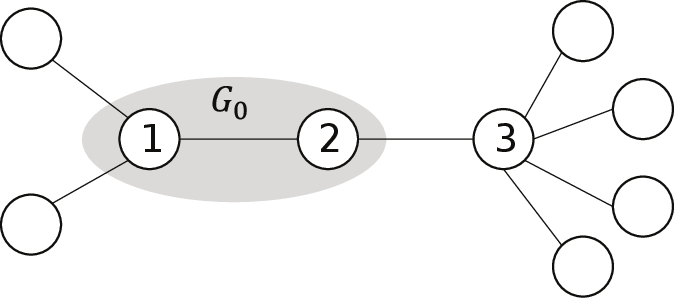}
            \caption{An example illustrating that querying a high degree node is not always beneficial given an initial subgraph $G_0$ when $T=k=1$ and $G_\text{gen-prun}^{(0)}$ correctly reflects the underlying graph $G$.}
            \label{fig:nodeselect}
    \end{center}
\end{figure}

However, such a selection strategy that chooses nodes whose residual degrees are high may not always  be beneficial as depicted in Fig. \ref{fig:nodeselect}. In this example, given an initial subgraph, querying node $v_1$ ($r_1^{(0)} = 2$) leads to an explored subgraph consisting of four nodes and three edges, while choosing node $v_2$ ($r_2^{(0)} = 1$) to query only discovers node $v_3$ and one edge incident to $v_3$. Since it is certain that $v_3$ is the most influential node, it should be contained in the explored graph for the seed node selection process. This observation motivates us to make use of the {\em geodesic distance} to potentially influential nodes, which represents the length in terms of the number of edges on the shortest path between two vertices, as a vital factor in the query node selection process. For example, in Fig. \ref{fig:nodeselect}, if we choose one of two nodes $v_1$ and $v_2$ such that the shortest path to $v_3$ yields a shorter geodesic distance as a queried node, then $v_2$ should be selected.

In this context, we want to  query each node by taking into account both 1) the residual degree (as high as possible) and 2) the sum of the shortest paths to potentially influential nodes (as low as possible). To this end, starting from the reinforced weighted graph  $G^{(t)}_\text{gen-prun}$, we first execute an IM algorithm to find the set of $k$ potentially influential nodes at the $t$-th query step, denoted as $\hat{S}^{(t)} \subseteq \mathcal{V}$.\footnote{In our work, degree discount heuristics \cite{degreediscount} are adopted to find $\hat{S}^{(t)}$ for computational efficiency.} Then, if a node $v \in \hat{S}^{(t)}$ belongs to the current explored subgraph ({\it i.e.}, $v \in \mathcal{V}_t$), then it is excluded from $\hat{S}^{(t)}$ ({\it i.e.}, $\hat{S}^{(t)} \leftarrow \hat{S}^{(t)}\setminus\{v\} $) so that only the potentially influential nodes in the unexplored part are possible candidates to be queried. Using the set $\hat{S}^{(t)}$, we now propose our own {\em topology-aware} ranking strategy by defining a ranking measure:

\begin{align}
\label{eq:ranking}
\text{rank}^{(t)}(u) = r_u^{(t)} -\alpha\sum_{v \in \hat{S}^{(t)}}\text{GD}^{(t)}(u,v),
\end{align}
where $\text{GD}^{(t)}(u,v)$ indicates the geodesic distance between two nodes $u$ and $v$, and $\alpha\ge0$ is a hyperparameter balancing two terms in (\ref{eq:ranking}). The impact of $\alpha$ is empirically validated in Section \ref{sec:5c}. The node whose $\text{rank}^{(t)}(u)$ is the highest among $u\in\mathcal{V}_t$ is finally chosen as the queried node in each query step. Intuitively, the potentially influential nodes serve as anchors providing a guidance on choosing the query node $u$ in the sense of minimizing the sum of geodesic distances between $u$ and nodes in $\hat{S}^{(t)}$, $\sum_{v \in \hat{S}^{(t)}}\text{GD}^{(t)}(u,v)$ in (\ref{eq:ranking}). It is worth noting that the number of seed nodes, $k$, is used not only in the SSP but also in the NDP.

We shall empirically validate the effectiveness of our query node selection using the topology-aware ranking strategy in Section \ref{sec:5}.

\subsection{Seed Set Selection}\label{sec:4b}

\begin{algorithm}[t]
\DontPrintSemicolon
\KwIn{$G^{(T)}_\text{gen-prun},k$}
\KwOut{$S$}
\SetKwBlock{Begin}{function}{end function}
\Begin($\textsf{ModifiedGreedy}$)
{
  $S \leftarrow \phi$\;
   \For{$i$ {\upshape \bf from} $0$ {\upshape \bf to} $k$}
  {
  \For{{\upshape \bf each} $v \in \mathcal{V}_T \backslash S$}
  {
  $s_v \leftarrow \hat{\sigma}(S \cup \{v\})$\;
  }
  $S \leftarrow S \cup \argmax_{v\in\mathcal{V}_T}\{s_v\}$\;
  }
  \Return{$S$}
}
\caption{\textsf{ModifiedGreedy}}\label{al2}

\end{algorithm}

We now turn to stating the SSP, which can be viewed as a generalization of the greedy IM algorithm \cite{IMorigin} and is described in Algorithm 2. First, the seed set $S$ is empty (see line 2). Then, we gradually find and add one node maximizing the so-called ``marginal expected spread'' to $S$, under the constraint that seed nodes can only be selected from the set of explored nodes,  $\mathcal{V}_T$ in the graph $G_\text{gen-prun}^{(T)}$ (see lines 3--6). For each candidate node $s_v$, we compute the marginal expected spread $\hat{\sigma}(S \cup \{v\})$ based on $G^{(T)}_\text{gen-prun}$ (see line 5).  We note that $\hat{\sigma}(\cdot)$ can be approximated via many approaches such as Monte-Carlo simulations, lazy update \cite{celf}, or reverse reachable sets \cite{IMM}. The node $s_v$ is added to $S$ if its corresponding $\hat{\sigma}(\cdot)$ is the largest (see line 6).

\subsection{Theoretical Analyses}\label{sec:4c}

First, we analyze the computational complexity of our \textsf{IM-META} method  as follows.

\begin{theorem}
The computational complexity of the proposed \textsf{IM-META} method is no higher than that of the greedy IM algorithm.
\end{theorem}

It is not difficult to show that the  complexity for the NDP is dominated by the network inference step, whose complexity is given by $\mathcal{O}(n^2)$, where $n$ is the number of nodes in $\mathcal{V}$, conditioned that the size of model parameters in MLPs does not scale with $n$. Of course, the computational complexity for this phase can be greatly reduced when  parallelization is applied. Thus, the bottleneck of \textsf{IM-META} is the greedy IM algorithm in the SSP, whose complexity is bounded by $\mathcal{O}(Rknm)$ (refer to \cite{celf} for more details), where $R$ represents the number of Monte-Carlo simulations to estimate the expected influence spread and $m$ is the number of edges in $\mathcal{E}$. As $R$ and $k$ are regarded as constants and $n < m$, the total complexity of \textsf{IM-META} is  given by $\mathcal{O}(nm)$, which lies in the same line as that of the greedy IM algorithm on the underlying graph $G$.

Next, we theoretically analyze the performance of our \textsf{IM-META} method.  More specifically, we show performance guarantees within the factor of $1-\frac{1}{e}$ of the optimum since a modified greedy algorithm is employed in the SSP of \textsf{IM-META}.

\begin{theorem}
Suppose that the NDP is completed, thus leading to the discovered subgraph $G_T$. Then, the proposed \textsf{IM-META} method, employing a greedy seed selection approach, achieves a $(1 -\frac{1}{e})$-approximation of the optimal solution with the optimal seed set $S^*$.
\end{theorem}

\begin{proof}

Let us recall an estimate of the expected influence spread, $\hat{\sigma}(S)$, used for the objective function in (2). Then, it is not difficult to show that the objective function $\hat{\sigma}(S)$ is monotone and sub-modular with respect to the seed set $S$ as long as the SSP is only concerned after the NDP is completed. Hence, following \cite{submodular}, given the function $\hat{\sigma}(\cdot)$, the greedy seed selection algorithm give a $(1-\frac{1}{e})$-approximation to the optimal solution with $S^*$ under our setting with network uncertainty. This completes the proof of Theorem 2.

\end{proof}

\section{Experimental Evaluation}\label{sec:5}

\subsection{Benchmark Methods}\label{sec:4b}

In this subsection, we present two baseline methods and one state-of-the-art IM approach against which we compare \textsf{IM-META}.

{\bf Random (Rand).} We randomly select a node to query for each round of exploration. Then, after $T$ query rounds, the greedy IM algorithm in \cite{IMorigin} is used to obtain $k$ seed nodes from the explored graph $G_T$ without network inference.

{\bf Depth-first search (DFS).} We randomly select a node to query out of all neighbors
 of the queried node in the previous round. Similarly as in Rand, the greedy IM algorithm is used to obtain $k$ seed nodes from the explored graph $G_T$.

{\bf CHANGE}. To facilitate an effective HIV intervention campaign, a state-of-art method was recently proposed in~\cite{c3}. This model was inspired by the friendship paradox, which states that the expected degree of a random node’s neighbor(s) is larger than that of the random node itself \cite{friendparadox}. In this context, a random neighbor of a node in $G_t$ is selected to query at each query step $t$ in the NDP. The greedy IM algorithm is also used after $T$ query rounds. Although CHANGE deals withs an end-to-end IM procedure, we still term the friendship paradox sampling as CHANGE for consistency with prior studies \cite{EIM2, EIM3, EIM4}.

Furthermore, for comparison with the fundamental limit of IM in our setting, we evaluate the performance of another case, termed {\bf Upper-Greedy}, using the greedy IM algorithm when the underlying graph $G$ is assumed to be {\it fully available}.

\subsection{Datasets}\label{sec:5a}
Four real-world datasets across several domains ({\it e.g.}, social and co-authorship networks with node metadata) commonly adopted in the literature are used as the underlying graph $G$, namely {\bf Coauthorship-CS (CS)}, {\bf Coauthorship-Physics (Physics)},\footnote{The two co-authorship networks are publicly available online (https://github.com/shchur/gnn-benchmark/tree/master/data).} {\bf Douban},\footnote{The dataset is publicly available online (https://github.com/maffia92/FINAL-network-alignment-KDD16).} and {\bf Ego-Facebook (Ego-FB)}~\cite{snapdata}. The four datasets are intentionally chosen from our empirical study to display different behaviors of \textsf{IM-META} and benchmark methods under various settings. 

We also make use of the associated node features provided in these datasets as the collected node metadata $\mathcal{X}$. For all experiments, we treat graphs as undirected. 
The statistics of each dataset, including the number of nodes and the number of edges, and the number of node features, are described in Table~\ref{tab:dataset}. Note that all metadata presented in these datasets are anonymized as binary values for privacy preservation.

\begin{table}[t]
\caption{Statistics of the four datasets, where $n$, $m$, and $|\mathcal{X}|$ denote the numbers of nodes, edges, and features of each node, respectively.}
\label{tab:dataset}
\centering
\begin{tabular}{llll}
\hline
Name                                             & $n$      & $m$  & $|\mathcal{X}|$      \\ \hline
Coauthor-CS & 18,333 & 34,493    & 6,805                                                                             \\ 
Coauthor-Physics & 168,788 & 495,924    & 8,415                                                                     \\ 
Douban                  & 3,906 & 8,164     & 538                                                                         \\ 
Ego-Facebook           & 534 & 4,813    & 262       \\ \hline
\end{tabular}
\end{table}

\subsection{Experimental Setup}\label{sec:setup}

In our experiments, the induced subgraph $G_0$ with four nodes is initialized, similarly as in \cite{EIM2}. For each evaluation, we run experiments over 10  different initialization of $G_0$  to compute the average score. The diffusion probability for the IC model is set to 0.1, which is the typical value from the literature (see \cite{IMorigin,celf} and references therein). As for the function \textsf{NetInference}, the dimension of embedding vectors is set to 256, and the Siamese neural network  model is trained via the Adam optimizer \cite{adamopt} with a learning rate of 0.1 and a batch size of 16 for all datasets. 

For each dataset, we randomly split the set of nodes into training/validation sets with a ratio of 90/10\%. The parameter $\alpha$ in (\ref{eq:ranking}) for designing our topology-aware ranking strategy and the threshold $\epsilon$ in the reinforced weighted graph generation are set to 1 and 0.5, respectively, unless otherwise stated.  For the function \textsf{ModifiedGreedy}, we employ the CELF algorithm \cite{celf} to ensure the correctness of our validation due to its theoretical performance guarantee, in which the number of Monte-Carlo simulations is set to 20,000. However, since the results of all competing methods would take too long to acquire when performing experiments on two large datasets, including Coauthor-CS and Coauthor-Physics,  when using CELF, we down-sample the size of both networks to 3,000 nodes using forest fire sampling unless otherwise stated, which is capable of preserving the degree distribution of the original graph \cite{graphsampling}. We note that the original CS and Physics datasets  without sampling are used to validate the scalability of \textsf{IM-META} in Section \ref{sec:546}.

All experiments are carried out with Intel (R) 12-Core (TM) i7-10700F CPUs @ 2.90 GHz and 16GB RAM. 

\subsection{Experimental Results}\label{sec:5c}
Our empirical study in this subsection is designed to answer the following key research questions (RQs):
\begin{itemize}
\item[$RQ1$] What is the growth of subgraphs in the query process of \textsf{IM-META} in comparison to the benchmarks?
\item[$RQ2$] What is the performance contribution of the network inference and query node selection modules?
\item[$RQ3$] How much does the \textsf{IM-META} method improve the performance of IM in terms of the expected influence spread over the benchmark methods?
\item[$RQ4$] How robust is \textsf{IM-META} in more challenging settings where a portion of node metadata are missing? 
\item[$RQ5$] What is the effect of model parameters on the performance of  \textsf{IM-META}?
\item[$RQ6$] How scalable is \textsf{IM-META} with growing size of the networks?
\end{itemize}

\subsubsection{Growth of Explored Subgraphs (RQ1)}\label{sec:541}

Given a small budget of node queries, it would be desirable to acquire explored subgraphs that are as large as possible since the size of subgraphs primarily determines the performance of IM. To comprehensively understand the query process, we are interested in examining how the explored subgraph $G_t$ scales with the number of queried nodes, $t$, on the four real-world datasets. To this end, we empirically assess the growth of $G_t$ resulting from four methods including \textsf{IM-META}, Rand, DFS, and CHANGE for all datasets, where the number of seed nodes, $k$, is set to 5 and the IC model is applied. Following the results in Fig. \ref{fig:size}, our findings can be summarized as follows:

\begin{figure}[t]
    \begin{center}
\vspace{0.3cm}
            \includegraphics[width=\linewidth]{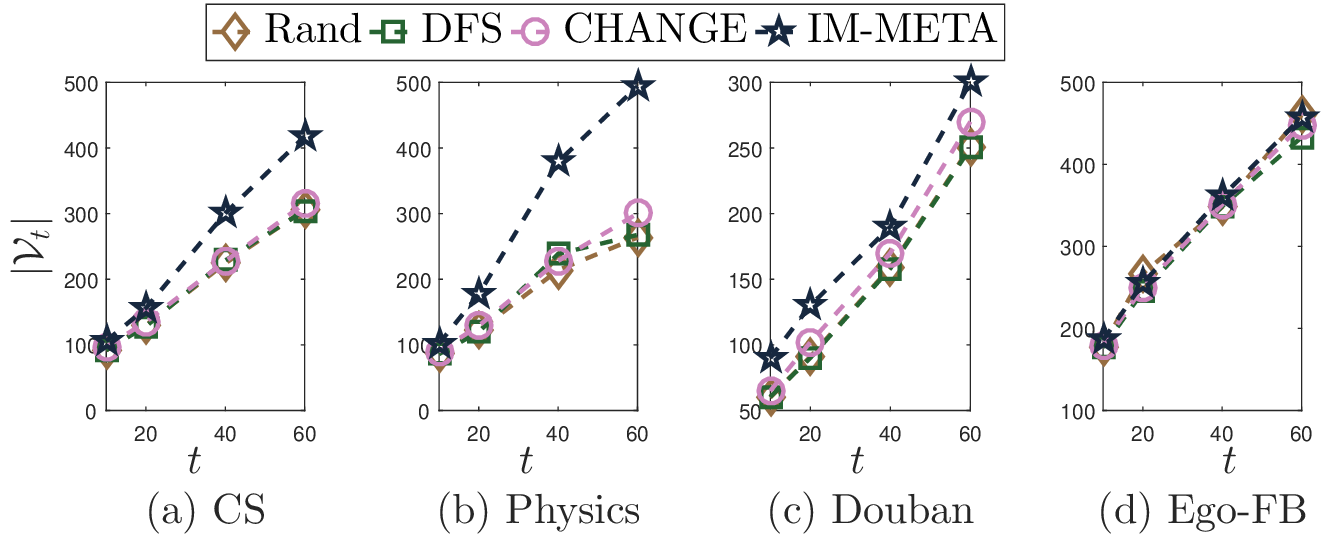}
            \caption{The size of explored subgraphs, $|\mathcal{V}_t|$, as a function of the number of queried nodes, $t$.}
            \label{fig:size}
    \end{center}
\end{figure}

\begin{itemize}
\item The size of each subgraph $G_t$ in the CS, Physics, and Douban datasets is relatively small compared to that of the underlying network $G$. For example, from Table \ref{tab:dataset} and Fig. \ref{fig:size}c, it is evident that $|\mathcal{V}_t|<0.1|\mathcal{V}|$  for the Douban dataset when $t=60$.

\item On the other hand, the Ego-FB network behaves differently from the other three datasets in such a way that $|\mathcal{V}_t|$ obtained by all methods is close to $|\mathcal{V}|$ even with querying only 60 nodes, which accounts for about 11.2\% of the total number of nodes in the Ego-FB  network. This is due to the fact that the diameter of ego-networks (created via breadth-first search (BFS) sampling) is inherently small; thus, it is capable of exploring a large number of neighbors by querying even a randomly selected node. 

\item For the CS and Physics datasets, the subgraph growth rate of \textsf{IM-META} is much faster than that of the other methods. This implies that the accuracy of network inference is sufficiently high due to the rich feature information of these two datasets (refer to Table \ref{tab:dataset}). 

\item In contrast, no such tendency is observed for the Douban and Ego-FB datasets. For the Douban dataset, the relatively low accuracy of network inference achieved by \textsf{IM-META} stems from the fact that node features are heavily anonymized. For the Ego-FB dataset, there is only a negligible gap in the subgraph size $|\mathcal{V}_t|$ among the four methods since querying only a small number of nodes leads to a sizeable exploration of subgraphs.
\end{itemize}

\subsubsection{Effectiveness of Each Component (RQ2)}
In order to discover what role each component plays in the success of the proposed \textsf{IM-META} method, we compare the performance of the original \textsf{IM-META} with its four variants.
\begin{itemize}
\item \textsf{IM-META(LR)}, \textsf{IM-META(GCN)}, and \textsf{IM-META(Transformer)}: For network inference, the Siamese neural network model is replaced by either logistic regression (LR), a graph convolutional network (GCN) \cite{gcn}, or a graph transformer block (Transformer) \cite{graphtransformer}. Specifically, for \textsf{IM-META(LR)}, we concatenate the two node feature vectors  ${\bf x}_u$ and ${\bf x}_v$ and adopt a simple MLP architecture in the hidden layer of size 512. For \textsf{IM-META(GCN)} and \textsf{IM-META(Transformer)}, we employ a single GCN layer and a graph transfomer block to train on a graph constructed from the explored subgraph $G_t$ and all other unexplored nodes, respectively.
\item \textsf{IM-META(Katz)} and \textsf{IM-META(degree)}: For query node selection, the topology-aware ranking strategy is replaced by the ranking based on two centrality measures including the Katz centrality and the degree centrality. Precisely, we compute the two centrality measures of all nodes in $\mathcal{V}_t$ using the inferred topology in $G^{(t)}_\text{gen-prun}$. Then, all nodes are ranked in descending order of the centrality measure. The node with the highest rank is finally chosen as the queried node in each query step.
\end{itemize}

The performance of \textsf{IM-META} and its above-mentioned variants is summarized in Table \ref{tab:abb}, in which we only show the results for the CS dataset when $k=5$ and the IC model is employed since the results for other datasets exhibit similar trends. Our findings are as follows.
\begin{itemize}
\item The original \textsf{IM-META} method always exhibits substantial gains over its variants, which demonstrates that each of our network inference and query node selection modules plays a crucial role in IM.

\item The performance of both \textsf{IM-META(GCN)} and \textsf{IM-META(Transformer)} is rather inferior to that of \textsf{IM-META(LR)} when the number of queried nodes is low due to a large number of unexplored edges. This implies that the deep learning-based models used for network inference exhibit overfitting and do not effectively deal with the highly uncertain network structure.

\item The centrality measures are not effective for node queries since they do not take the strength of influence into account.
\end{itemize}

\begin{table}[]

\caption{The performance comparison among \textsf{IM-META} and its variants on the CS dataset in terms of the expected influence spread $\sigma$ when $k=5$ and the IC model is applied. Here, the best performers are highlighted by bold fonts.}
\label{tab:abb}
\centering
\begin{tabular}{lllll@{}}

\toprule
\multirow{2}{*}{Method}          & \multicolumn{4}{l}{Number of queried nodes}                   \\ \cmidrule(l){2-5} 
                                 & 10            & 20            & 40            & 60            \\ \cmidrule(r){1-5}
\textbf{\textsf{IM-META}} & \textbf{45.1} & \textbf{47.1} & \textbf{50.9} & \textbf{61.5} \\
\textsf{IM-META(LR)} & 39.9          & 41.8          & 42.3          & 47.4          \\
\textsf{IM-META(GCN)}                 & 38.6          & 41.3          & 42.5          & 48.8          \\ 
\textsf{IM-META(Transformer)}                 & 38.4          & 41.2          & 42.2          & 48.5          \\ 
\textsf{IM-META(Katz)}                 & 42.6          & 46.2          & 48.9         & 55.5          \\ 
\textsf{IM-META(degree)}                 & 41.9          & 46.3          & 48.8          & 59.2          \\ 
\bottomrule
\end{tabular}
\end{table}

\subsubsection{Comparison With Benchmark Methods (RQ3)} 

\begin{figure}[t]
    \begin{center}
\vspace{0.3cm}
            \includegraphics[width=\linewidth]{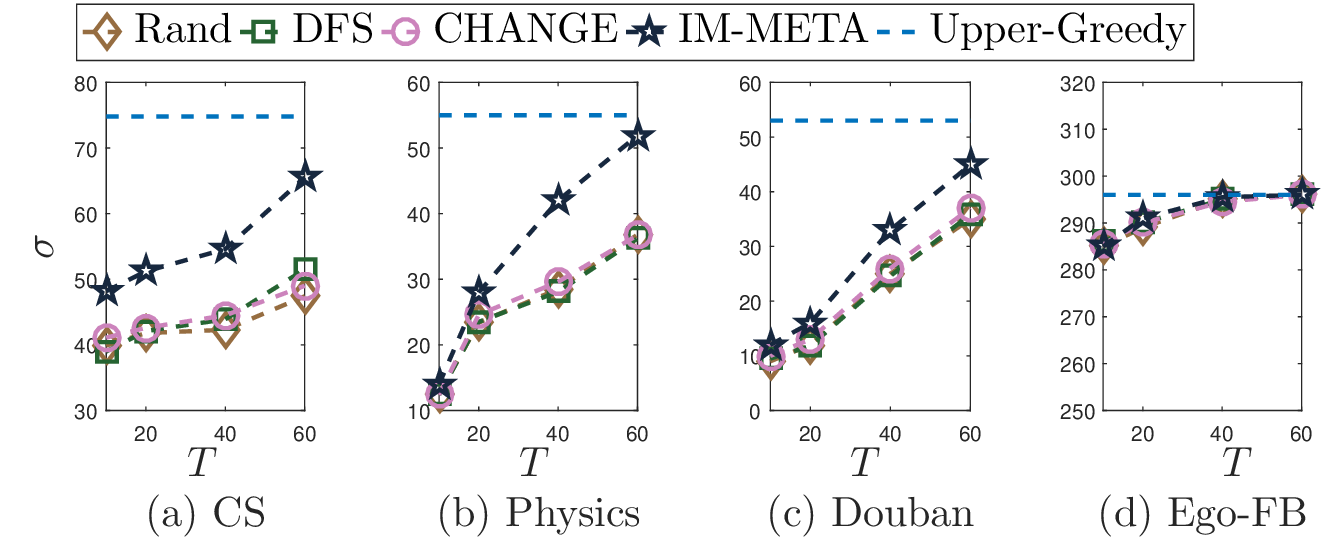}
            \caption{$\sigma$ as a function of $T$ when $k = 5$ (IC model).}
            \label{fig:5}
    \end{center}
\end{figure}

\begin{figure}[t]
\vspace{0.3cm}
    \begin{center}
            \includegraphics[width=\linewidth]{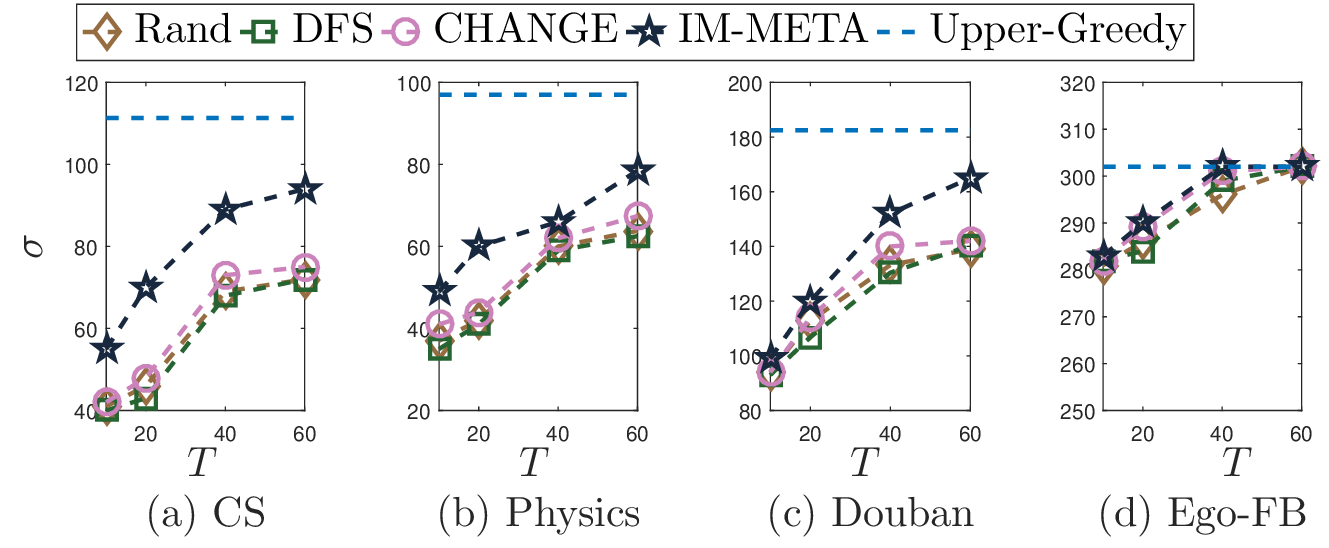}
            \caption{$\sigma$ as a function of $T$ when $k = 10$ (IC model).}
            \label{fig:6}
    \end{center}
\end{figure}

\begin{figure}[t]
\vspace{0.3cm}
    \begin{center}
            \includegraphics[width=\linewidth]{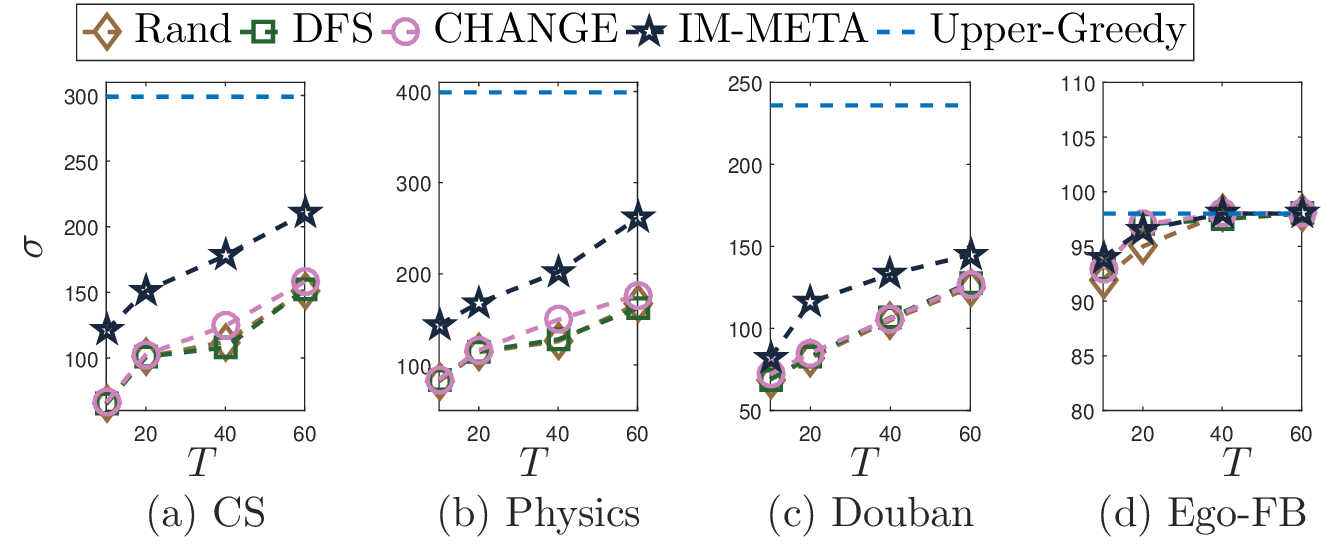}
            \caption{$\sigma$ as a function of $T$ when $k = 5$ (WC model).}
            \label{fig:7}
    \end{center}
\end{figure}

\begin{figure}[t]
\vspace{0.3cm}
    \begin{center}
            \includegraphics[width=\linewidth]{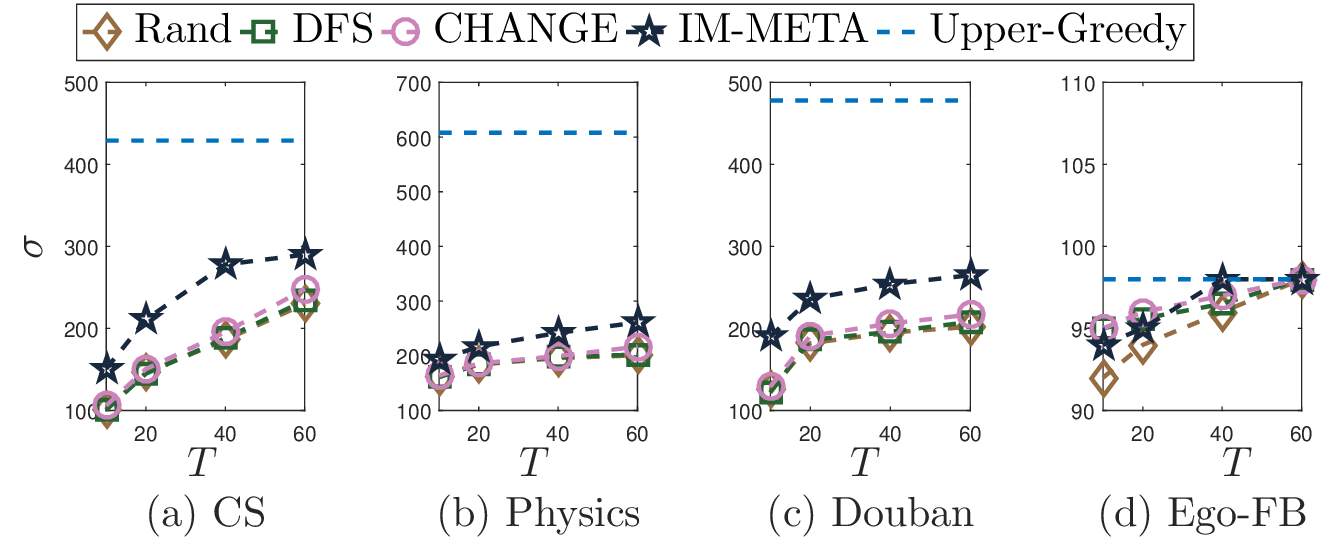}
            \caption{$\sigma$ as a function of $T$ when $k = 10$ (WC model).}
            \label{fig:8}
    \end{center}
\end{figure}

To comprehensively compare the performance of \textsf{IM-META} to the benchmark methods and Upper-Greedy, we consider the expected influence spread $\sigma(S)$ for a set of selected seed nodes, $S$,  using all four real-world datasets.\footnote{To simplify notations, $\sigma(S)$ will be written as $\sigma$ unless otherwise specified.} Figs. \ref{fig:5}--\ref{fig:8} illustrate $\sigma$ versus the number of node queries, $T$, where $k=\{5,10\}$ and both IC and WC models are employed. We make the following insightful observations:
\begin{itemize}
\item \textsf{IM-META} consistently and remarkably outperforms all benchmark methods except for the Ego-FB dataset.
\item For the Ego-FB dataset, all methods achieve comparable performance to that of Upper-Greedy even with a very small number of node queries, {\it e.g.},  $T=10$. This is not surprising due to the enormous size of explored subgraphs as discussed in Section \ref{sec:541}, which results in the inclusion of (almost) all influential nodes in any explored subgraph.
\item In comparing the IC and WC models, it is likely that, for the WC model, the performance gap between the \textsf{IM-META} and other methods is higher. This is because \textsf{IM-META} is capable of estimating the expected degree of nodes; thus, the diffusion probability can be more precisely represented in the WC model by employing such methods that make use of node features.
\item The performance of \textsf{IM-META} tends to be closer to that of Upper-Greedy when the number of seed nodes, $k$, is relatively smaller. As an example, in the Physics dataset, \textsf{IM-META} achieves about 93\% of the expected influence spread resulting from Upper-Greedy by querying only 60 nodes, which accounts for 5\% of the total number of nodes in the given network, when $k=5$ and the IC model is applied (see Fig. \ref{fig:5}b).

\end{itemize}
Overall, \textsf{IM-META} best approaches the upper bound of the greedy IM algorithm (which has access to the entire network).

\subsubsection{Robustness to the Degree of Available Node Metadata (RQ4)}
We now evaluate the performance of \textsf{IM-META} in a more difficult situation that often occurs in real environments, where only a portion of node metadata in the set $\mathcal{X}$ are collectible  (available). To this end, we randomly remove \{20, 40, 60, 80\}\% of node features associated with each node in a given dataset. In these experiments, we show only the results for the setting where $T$ and $k$ are set to 60 and 5, respectively. We omit results for the WC model since they follow similar trends. 

\begin{figure}[t]
\vspace{0.3cm}
    \begin{center}
            \includegraphics[width=\linewidth]{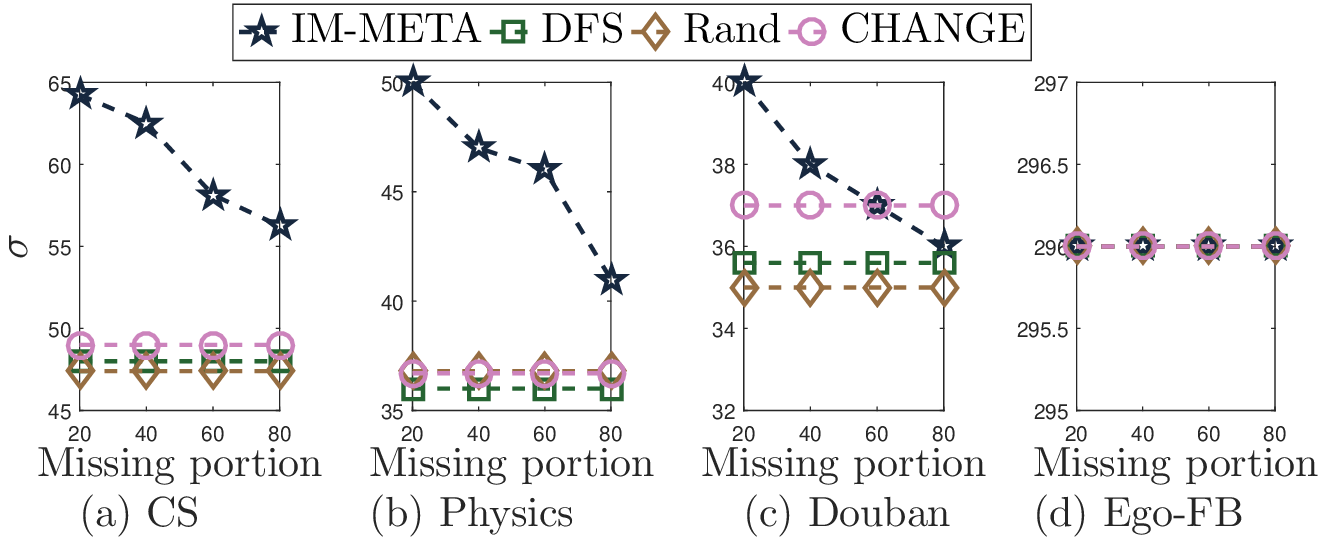}
            \caption{$\sigma$ according to the missing portion of collected metadata when $T$ and $k$ are set to 60 and 5, respectively (IC model).}
            \label{fig:9}
    \end{center}
\end{figure}

The performance comparison between \textsf{IM-META} and the three benchmark methods for all datasets is presented in Fig. \ref{fig:9}. One can see that the performance of \textsf{IM-META} gradually degrades when more node features are removed from $\mathcal{X}$. Our findings for the CS and Physics datasets demonstrate that, in fringe scenarios, {\it e.g.}, the case where 80\% of node features are removed, \textsf{IM-META} still exhibits clear gains over all benchmark methods. This implies that, even under the circumstance where a number of node features are missing, the \textsf{IM-META} method is capable of correctly adding inferred edges in the subgraph exploration process, which enables us to better select the set of influential seed nodes. For the Douban dataset with 80\% removal of node features, \textsf{IM-META} is slightly inferior to CHANGE yet still reveals  gains over other two methods. On the other hand, for the Ego-FB dataset, all methods achieve the performance of Upper-Greedy, meaning that the degree of available  node metadata does not affect any performance degradation.

\subsubsection{Hyperparameter Sensitivity (RQ5)}\label{sec:544}

\begin{figure}[t]
\vspace{0.3cm}
    \begin{center}
            \includegraphics[width=0.85\linewidth]{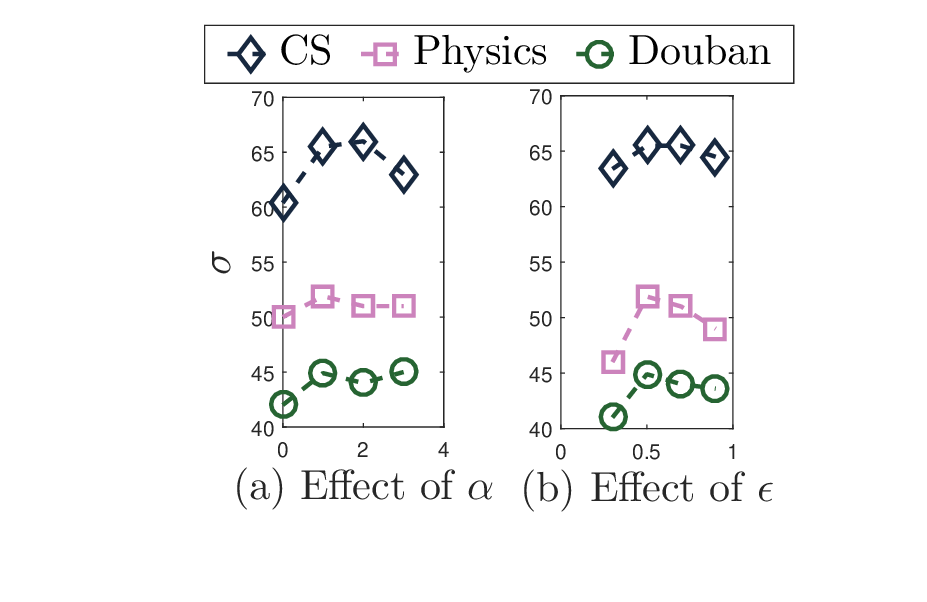}
            \caption{The expected influence spread $\sigma$ achieved by \textsf{IM-META} according to different values of hyperparameters $\alpha$ and $\epsilon$ when $T$ and $k$ are set to 60 and 5, respectively, and the IC model is applied.}
            \label{fig:sens}
    \end{center}
\end{figure}
In addition to four key RQs that we answered, another important question concerns the importance of tuning hyperparameters. To answer the question, we carry out experiments as follows.

In Fig. \ref{fig:sens}, we investigate the impact of hyperparameters $\alpha$ and $\epsilon$ on the performance of \textsf{IM-META} when $T=60$ and $k=5$. Recall that $\alpha$ adjusts the balance between two terms for the query node selection in (\ref{eq:ranking}), {\it i.e.}, the residual degree and the sum of the shortest paths to potentially influential nodes, and $\epsilon$ is the threshold controlling the number of {\it confident} edges for the reinforced weighted graph generation. When a hyperparameter ($\alpha$ or $\epsilon$) varies so that its effect is clearly revealed, the other parameter is set to the pivot values $\alpha = 1$ and $\epsilon = 0.5$. In these experiments, we only show the results from the IC model since those from the WC model follow a similar trend. Moreover, experimental results for the Ego-FB dataset are excluded since all combinations of hyperparameters return the same performance on this dataset. We observe the following:

\begin{itemize}
\item The values of $\alpha$ and $\epsilon$ for which the highest $\sigma$ is achieved tend to be almost identical for all datasets. Specifically, we find that the case of $(\alpha,\epsilon)=(1,0.5)$ exhibits the best performance almost consistently.
\item We investigate the case of $\alpha=0$ where the residual degree is only taken into account in our topology-aware ranking strategy. In this case, the lowest performance is achieved over all $\alpha$'s regardless of datasets. This finding validates the necessity of the {\em shortest path to potentially influential nodes} in judiciously choosing queried nodes by setting the value of $\alpha$ appropriately.
\item We next examine the impact of $\epsilon$ on the performance. Setting $\epsilon$ to a high value ({\it e.g.}, $\alpha=0.9$) leads to relatively low performance for all datasets. This setting is indeed undesirable since it makes our reinforced weighted graph $G_\text{gen-prun}^{(t)}$ very sparse with only a small number of edges. Hence, it is crucial to suitably determine the value of $\epsilon$ in guaranteeing satisfactory performance.
\end{itemize}
Overall, the results of our parameter sensitivity tests are as expected. The importance of the topology-aware ranking strategy, leveraging the residual degree and path information, for IM is clear from our empirical findings; and appropriately deciding the density of the generated weighted graphs is beneficial for the selection of seed nodes.

\subsubsection{Scalability (RQ6)}\label{sec:546}

\begin{figure}[t]
\vspace{0.3cm}
    \begin{center}
            \includegraphics[width=0.85\linewidth]{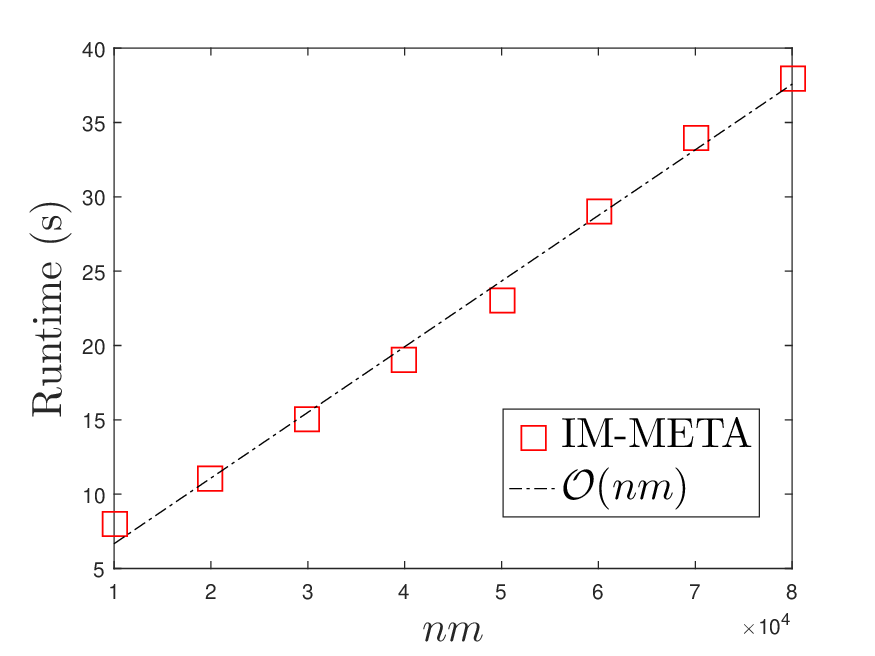}
            \caption{The runtime complexity of \textsf{IM-META} with respect to different network size products $nm$.}
            \label{fig:complex}
    \end{center}
\end{figure}

To empirically validate the computational complexity of \textsf{IM-META} shown in Section 4.3, we conduct experiments using a set of graphs synthetically generated from the Erdős–Rényi graph model [48], where $(n,m)$ = \{(50, 200), (50, 400), (50, 600), (100, 400), (100, 500), (100, 600), (100, 700), (200, 400)\}. The metadata associated with each node are generated as a 10-dimensional all-ones vector. The number of Monte-Carlo simulations, the number of seed nodes, and the number of queries are set to 20,000, 5, and 10, respectively. We also set the number of edges to be inferred identically to the number of edges in the underlying network. In Fig. \ref{fig:complex}, we show the measured runtime complexity of \textsf{IM-META} (in seconds) with respect to different network size products $nm$ according to the above-mentioned setting. From the asymptotic fit of $\mathcal{O}(nm)$ with an appropriate bias (dashed line in the figure), we can clearly see that the computational complexity scales as  $\mathcal{O}(nm)$, which validates our analysis in Theorem 1.

Furthermore, to verify that \textsf{IM-META} is applicable in real-world networks whose sizes are typically large, we measure its runtime complexity using the original Coauthor-CS and original Coauthor-Physics datasets without sampling.  For the function \textsf{ModifiedGreedy}, we employ the SSA algorithm \cite{SSA} for computational efficiency. It is worth noting that we can also replace SSA with any heuristics for much faster runtime as the seed node selection is model-agnostic. Table \ref{tab:SSA2} shows the measured runtime of the NDP and SSP each in \textsf{IM-META} as well as \textsf{IM-META} itself (in seconds) with respect to the number of node queries, $T$, where $k$ is set to 10 and the IC model is employed. 
Empirical evidence reveals that the runtime of the NDP is far smaller than that of the SSP. Moreover, the computational complexity of the SSP, which is the bottleneck, does not scale in proportion to $T$, leading to a gradual rise in the runtime of \textsf{IM-META}.

\begin{table}[]

\caption{The runtime complexity of the NDP and SSP each in \textsf{IM-META} as well as \textsf{IM-META} itself with respect to the number of node queries, $T$, for $k=10$ on two large-scale networks.}
\label{tab:SSA2}
\centering
\begin{tabular}{lllll@{}}

\toprule
\multirow{2}{*}{Runtime (s)}          & \multicolumn{4}{l}{Number of queried nodes}                   \\ \cmidrule(l){2-5} 
                                 & 100            & 200            & 300            & 400            \\ \cmidrule(r){1-5}
NDP  in \textsf{IM-META}  & 144          & 221          & 233          & 245          \\
SSP  in \textsf{IM-META}                 & 1,454          & 1,454          & 1,454          & 1,454          \\ 
{\bf \textsf{IM-META} (Total)}                 & {\bf 1,598} & {\bf 1,675}          & {\bf 1,687}          & {\bf 1,699}           \\ 
\bottomrule
\end{tabular}
\end{table}

\section{Concluding Remarks}\label{sec:6}
In this paper, we investigated the impact and benefits of node metadata in aiding solutions to the IM problem in topologically unknown networks, which was an open and challenging problem due to the noisy nature of metadata and uncertainties in connectivity inference. To tackle this challenge, we introduced \textsf{IM-META}, a novel end-to-end solution that aims to find both a set of seed nodes and another set of nodes to query by making full use of the node metadata. Specifically, we developed the NDP that iteratively performs three steps: 1) learning the relationship between collected metadata and edges via a Siamese neural network model, 2) constructing a reinforced weighted graph by selecting only a limited number of {\it confident} edges, and 3) discovering the next node to query using the proposed topology-aware ranking strategy balancing between degree centrality of a target node and its geodesic distance to potential seeds. We then employed a modified IM algorithm to find the seed set, given the ultimately explored subgraph. Using four real-world datasets, we demonstrated that, for almost all cases, the proposed \textsf{IM-META} method consistently outperforms benchmark IM methods in terms of the expected influence spread. The results of our evaluation showed that \textsf{IM-META} achieves up to 93\% of the performance of an upper-bound approach that has access to the full network, while IM-META queried only 5\% of the underlying network. We also found that \textsf{IM-META} is robust to a lack of available node metadata ({\it e.g.}, the performance even in extreme settings in which 80\% of metadata are missing) and various hyperparameter settings. Overall, our results showed that even very limited access to node metadata and restricted queries to the network structure are sufficient to perform effective influence maximization.

Potential avenues of future research include the design of self-supervised learning models to further enhance the accuracy of network inference based on the node metadata of unexplored nodes. Alternatively, reinforcement learning could be incorporated into the framework for query node selection to further boost the performance. It could also be of interest to theoretically analyzing the gap between the upper bound performance and our achievability as a function of the number of queried nodes. Furthermore, it is worth noting that previous endeavors addressing IM in temporal or dynamic networks have not focused on the effective exploration of subgraphs. Consequently, our \textsf{IM-META} can be accommodated as a possible extension, facilitating enhanced subgraph exploration in temporal or dynamic network scenarios.

\section*{Acknowledgments}
This work was supported by the National Research Foundation of Korea (NRF), Republic of Korea Grant by the Korean Government through MSIT under Grants 2021R1A2C3004345 and
RS-2023-00220762, by the Institute of Information and Communications Technology Planning and Evaluation (IITP), Republic of Korea Grant by the Korean Government through MSIT (6G
Post-MAC---POsitioning and Spectrum-Aware intelligenT MAC for Computing and Communication Convergence) under Grant 2021-0-00347, and the Postdoctoral Scholarship Programme of Vingroup Innovation Foundation (VINIF), code VINIF.2023.STS.58.

\bibliography{AllBib_clean} 
\bibliographystyle{IEEEtran}

\end{document}